\documentstyle[editedvolume,namedreferences]{crckapb} 

\begin{opening}
\title{A MODEL OF THE BROAD-BAND CONTINUUM OF NGC~5548}
\author{P. MAGDZIARZ}
\institute{Astronomical Observatory, Jagiellonian University\\
           Orla 171, 30-244 Cracow, Poland}
\author{O. BLAES}
\institute{Department of Physics, University of California\\
           Santa Barbara, CA 93106, USA}
\end{opening}
\begin{document}

\begin{abstract}
We discuss a model of the central source in Seyfert~1 galaxy NGC~5548. 
The model assumes a three phase disk structure consisting of a cold outer
disk, a hot central disk constituting a Comptonizing X/$\gamma$ source,
and an intermediate unstable and complex phase emitting a soft excess
component. The model qualitatively explains broad-band spectrum and
variability behavior assuming that the soft excess contributes
significantly to the continuum emission and drives variability by
geometrical changes of the intermediate disk zone. 
\end{abstract}

\section{Introduction}

Broad-band spectral analysis of NGC~5548 shows that a soft X-ray excess
dominating below 1 keV is related to the disk and may contribute
significantly to the source energetics (Magdziarz et al.\ 1997). Since an
optical/UV continuum requires the disk temperature of order a few eV, the
soft excess component needs an additional continuum emitting phase. The
standard opticly thick accretion disk (Shakura \& Sunyaev 1973) breaks
down at high accretion rate and, thus it puffs up producing the central,
hot ($\sim$100 keV) region (e.g.\ Shapiro, Lightman \& Eardley 1976) 
which constitutes the Comptonizing X/$\gamma$ source (e.g.\ Zdziarski et
al.\ 1997). In a such model the third intermediate phase of the disk (with
characteristic temperature $\sim$100 eV) naturally appears as an effect of
an instability. 

\section{Physics of the central source}

Correlated variability of the spectral index, amount of reflection, and
the total flux emitted in the X/$\gamma$ continuum (Magdziarz et al.\
1997) suggests that a number of seed photons controls the Compton cooling
of the central X/$\gamma$ source. This explains that the source is soft in
a bright state and hard (i.e.\ photon starved) in a faint state. The
opposite relation in the overall optical to soft excess continuum suggests
that the main contributor of seed photons for Comptonization is the soft
excess.  Lack of ionized Compton reflection and lack of emission lines in
EUV range requires complex structure of the soft excess. If the part of
emitting matter is Thomson thin and dense it may emit dominant part of
energy in a form of EUV pseudocontinuum (e.g.\ Kuncic, Celotti \& Rees
1997). 

\section{Phenomenological model}

The model assumes that the observed variability is driven by geometrical
changes of the unstable inner edge of the cold disk on thermal time scales
(cf.\ Kaastra \& Barr 1989; Loska \& Czerny 1997). This edge couples to
the hot plasma by energy reprocessing modulated with the solid angle of
cold matter seen from the central hot disk. The instability of the
intermediate zone between the cold and the hot phase puffs up the inner
cold disk and leads to its fragmentation which may be responsible for the
observed self-organized critically behavior of variability (e.g.\ Leighly
1997). The unstable zone has a complex structure of clouds or filaments
which explains the observed continuum of cold Comptonization in the soft
excess range (Magdziarz et al.\ 1997). The positive feedback between the
number of seed photons emitted from the unstable inner edge of the cold
disk and the energy radiated out from the hot central disk produces
substantial nonlinearity in the variability and naturally explains the
correlation of the total flux emitted in X/$\gamma$ continuum with the
spectral index and amount of reflection. In such a physical picture, the
soft excess dominantes energetics of the broad-band spectrum and drives
variability of the central source.

\end{document}